\documentstyle[12pt,epsfig]{article}

\newcommand{\be}{\begin{equation}}
\newcommand{\ee}{\end{equation}}

 1
\font\elevenrm=cmr10 scaled\magstep 1
 1

\begin{document}
\vspace*{1.8cm}
  \centerline{\bf WHAT IS THE STANDARD MODEL OF ELEMENTARY PARTICLES}
  \centerline{\bf AND WHY WE HAVE TO MODIFY IT\footnote{Presented 
at the Vulcano 2000 Workshop ``Frontier objects in astroparticle
and particle physics'', May 22-27, Vulcano, Italy.
In the spirit of the Workshop, I tryed to provide
an essential vocabulary and perhaps 
entry points for colleagues working 
in different fields of research.
I renounced to present a lists of 
references regarding experiments,
and I mostly limited the 
discussion to few ideas and 
theoretical leitmotifs; I apologize 
for the arbitrariness of the selection.}}
\vspace{1cm}
  \centerline{FRANCESCO VISSANI}
\vspace{1.4cm}
  \centerline{INFN, LABORATORI NAZIONALI DEL GRAN SASSO,}
  \centerline{\elevenrm I-67010, Assergi (AQ) Italy}
\vspace{3cm}
\begin{abstract}
We introduce the standard model 
of elementary particles and discuss the reasons why 
we have to modify it. 
Emphasis is put on the indications from the
neutrinos and on 
the role of the Higgs particle; 
some promising theoretical ideas, 
like quark-lepton symmetry, 
existence of super-heavy ``right-handed'' neutrinos,
grand unification and supersymmetry 
at the weak scale, are introduced 
and shortly discussed.
\end{abstract}
\vspace{2.0cm}

\section{The standard model of elementary particles}
In this section we define the standard 
model of elementary particles
(SM) by introducing the particle content
and the parameters of this model. 
We aim at giving a short introduction, 
at illustrating how this model is used, 
and at discussing (some of) its merits 
and limitations 
(to go deeper in the subjects touched, 
one could make reference 
to the seminal papers on the SM \cite{gws},
to books and reviews \cite{books,hist},
and to the {\sc Particle Data Group} biannual 
report \cite{pdg}).

\subsection{The basic blocks}

There are two groups of spin 1/2 particles, quark and leptons,
divided in 4 groups with different values of the electric charge $Q$
$$
\mbox{\sc MATTER FERMIONS}\ \ \
\left\{ 
\begin{array}{lllll}
\mbox{\bf leptons} & Q=0,&  \nu_{\rm e}\ \nu_\mu\ \nu_\tau ;&
Q=-1,& \rm{e}\ \mu\ \tau . \\
\mbox{\bf quarks} & Q=2/3,& u\  c\ t; & Q=-1/3,& d\ s\ b .
\end{array}
\right.
$$
(respectively named: e-, mu- and tau-neutrinos; electron, muon, tau;
up, charm and top quarks;  down, strange and bottom quarks;
recall the existence of their {\em anti}-particles, with opposite charge).
A different mass distinguishes among the 
three particles in each 
group\footnote{Incidentally, {\em mass} and {\em spin} 
are fundamental 
entities (=they identify certain irreducible representations)
of the underlying space-time group  of symmetry
(which has as subgroups the 4-dimensional 
translations and rotations--Lorentz subgroup). 
Note however that such an axiomatic 
definition of ``mass'' requires that the system 
can be considered isolated, that is 
usually true only in particular conditions.}
(neutrino masses are to a certain extent special, and we 
will discuss them in a few pages).
Thence, the fundamental particles can be 
arranged in three ``families'', with increasingly  
heavier members--but otherwise identical.
Free quarks have never been observed. 
It is thought that they can manifest as such 
only in very energetic processes; 
and that they necessarily bind to form the
``hadrons'' 
$$
\mbox{\sc HADRONS}\ \ \
\left\{
\begin{array}{ll}
\mbox{{\bf mesons} ($q\bar{q}$ states--bosons)} & \ \ \ \ \ \
\pi^\pm\ \pi^0\ K\ \rho\ ...\ \phi\ ...\ J/\Psi\ ...\\
\mbox{{\bf baryons} ($q q q$ states--fermions)} & \ \ \ \ \ \
p\ n\ \Delta^\pm\ ...\ \Lambda\ ...\ \Lambda_b\ ...
\end{array}
\right.
$$
the binding is provided by ``strong'' interactions, 
a prerogative of the quarks; the forces between nuclei 
are regarded as residual strong interactions.

Actually, the list of {\em stable} 
spin 1/2 particles is 
even shorten than the above ones; the electron, 
the proton $p$ (perhaps), the neutrinos (perhaps).

Then come the integer spin fundamental particles, that have 
the special role of {\em mediating} the interactions:
\begin{itemize}
\item The graviton (spin 2) related to \underline{gravitational} 
forces (that, strictly speaking are not part of the SM);
\item the gluon $g$ (spin 1) that carries \underline{strong} interactions;
\item the photon $\gamma$ (spin 1) that carries \underline{electromagnetic} 
interactions;
\item the $W^\pm $ and $Z^0$ 
bosons (spin 1) that originate \underline{charged and 
neutral} (current)
\underline{weak} interactions (responsible for 
instance of the nuclear $\beta$ decays--with 
emission of electrons);
\item and finally the {\sc Higgs} boson (a scalar, with spin zero)
that mediates \underline{{\sc Yukawa}} interactions (see below).
\end{itemize}
The last one is a hypothetical particle; 
however it has an outstanding importance in
the SM.
\subsection{How interactions are understood \label{sec:gp}}

There is a concept that 
underlies the theory 
of spin 1 interactions:
the gauge principle \cite{yangmills}
(which can be given
an essential role in 
formulating the theory of gravity, 
too; and perhaps, the one 
of {\sc Yukawa} 
interactions---but this is already
a speculation).
We will introduce it, 
starting to venture 
in the formalism.
Let us consider the equation 
that describes the free
propagation of a relativistic 
spin 1/2 particle:
\begin{equation}
\left[ i \!\!\! \sum_{\mu=0,1,2,3} \gamma^\mu \frac{\partial}{\partial x^\mu} 
-m \right] \psi(x)=0.\ \ \ \ \ \mbox{({\sc Dirac} equation)}
\end{equation}
Here, $m$ is the mass of the particle;
$\gamma^\mu$ are $4\times 4$ matrices that guarantee 
the covariance of the equation; 
$\psi$ has 4 components, necessary to 
describe the spin of a massive particle
(thenceforth named: 4-spinor). 
The spinorial index and $x$
describe the ``space-time'' 
transformation properties of $\psi;$
if we add more indices, we can construct
equations that are covariant under other 
symmetries\footnote{It is well known 
that symmetries have a central role 
in elementary physics, being related 
to conservation laws; 
the gauge principle further
augments their importance.}.
To be specific, let us consider 
the transformation $\psi_a\to U_{ab} \psi_b,$ where we 
parameterise $U$ by introducing the ``generators'' 
$T^B_{ab}$: $U=\exp( i \sum_B \alpha^B T^B)$ (the imaginary unity
$i$ in the equations above are just due to the tradition; it is important
instead to recall that the number of generators 
is a characteristic of the group of symmetry; 
1 for U(1), 3 for SU(2), 8 for SU(3), etc---SU($n$) 
being the group of
unitary matrices of 
dimension $n$ with unit determinant). 
The symmetry consist in the fact that 
the trasformed spinor  still obeys 
the {\sc Dirac} equation
(=the equation is ``covariant'' under the transformation);
and this is easy to prove, since neither the mass 
nor the $\gamma$-matrices are transformed.

However, in consideration 
of the local character of the space-time,  one is lead to 
wonder whether {\em local} symmetries 
$U(x)=\exp( i \sum_B \alpha^B(x) T^B)$ also hold true.
One readily verifies that the {\sc Dirac} equation
is not covariant in this enlarged context, 
due to the fact that the partial 
derivative $\partial/\partial x^\mu$ 
(that describes how the particle propagates 
in the space-time) {\em do act} on 
the parameters of the transformation. 
The ``gauge principle'' consists 
in insisting on covariance, by 
modifying the {\sc Dirac} equation in the following manner:
\begin{equation}
\begin{array}{l}
\displaystyle
\left[ i \!\!\! \sum_{\mu=0,1,2,3} \gamma^\mu 
\left(\frac{\partial}{\partial x^\mu}
+ i g A^B_\mu T^B\right) -m \right] \psi(x)=0 \\[3ex]
\displaystyle
A^B_\mu T^B \to U A^B_\mu T^B U^{-1} -\frac{i}{g}\ U 
\frac{\partial U^{-1}}{\partial x^\mu} 
\label{exex}
\end{array}
\end{equation}
we introduced a parameter (``coupling'') 
$g,$ and a set of fields labelled 
with space-time index $\mu$ 
(that transforms as a 4-vector=that 
describes a spin 1 particle);
the indices $a,b...$ are not indicated 
explicitly and a sum over $B$ is understood.
Different group of symmetries,
different particles. 
Note that in the special 
case of the U(1) symmetry group, only one particle $A_\mu$
must be introduced,
and due to the commutativity in the group, it has simple 
transformation properties since $U A_\mu T U^{-1}= A_\mu T.$
Identifying $g$ with the electron charge {\em e}, 
$T$ with value of the charge in unities of {\em e}, 
it becomes evident that the U(1) theory is just
the electromagnetism\footnote{Actually, 
this is the reason why 
one speaks of ``gauge principle'', since we rather directly 
generalise the well-known gauge invariance to induce 
the existence of new interactions related to new symmetries.}. 
For non-commutative groups
(say, SU($n$), where $U\cdot V\neq V \cdot U$ in general), 
instead, one notes that these spin-1 particles
transform in a non trivial manner {\em also} under global (non-local)
transformations, exactly due to the $U A^B_\mu T^B U^{-1}$ term.
Then, it comes without surprise 
the fact that these particle interact not only 
with the matter fermions (quarks and leptons) 
but also among themselves; however, 
this fact is of great significance,
since it means that the ``superposition principle'' of ordinary 
electromagnetic interactions (``the light does 
not interact with the light'') is not of general validity\footnote{This 
is among the most prominent peculiarity of the strong interactions, and 
a crucial ingredient to account for their behaviour. Note incidentally
that the existence of bound states of gluons (``glueballs'')
has been postulated, and there is theoretical support and 
circumstantial evidence of the correctness of this hypothesis, see
{\em e.g.}\ \cite{pdg}.}.

\subsection{First foundation of the standard model}
Now we can appreciate the first foundation of the standard model, 
namely, its gauge group 
$$
G_{321}=\mbox{SU(3)}_c\times\mbox{SU(2)}_L\times\mbox{U(1)}_Y
$$
the sign $\times$ means 
that the three subgroups commute (=they do not
communicate among them---neither they 
do, incidentally, with the space-time symmetries). 
The SU(3)$_c$ part is related 
to the existence of gluons, the rest (``electroweak'' group)
to the photon and the spin 1 bosons related to 
weak interactions.

\begin{table}[tb]
\begin{center}
\begin{tabular}{|c|ccr|}
\hline
& SU(3)$_c$ & SU(2)$_L$ & U(1)$_Y$ \\
\hline
$q_{L\alpha}$ & 3 & 2 & 1/6\ \\
$u_{R\alpha}$ & 3 & 1 & 2/3\ \\
$d_{R\alpha}$ & 3 & 1 & $-1/3$\ \\
$l_{L}$ & 1 & 2 & $-1/2$\ \\
$e_{R}$ & 1 & 1 & $-1$\ \\
\hline
\end{tabular}
\end{center}
\caption{Quark and leptons assignment in the SM.
$\alpha=1,2,3$ is the SU(3)$_c$ index. The SU(2)$_L$ ``doublets''
can be decomposed as: $q_L=(u_L,d_L)$ and 
$l_L=(\nu_{{\rm e}L},{\rm e}_L).$ 
Last column gives the value of $Y.$
This structure is just replied 
for the other two families of matter fermions.\label{tab:1}}
\end{table}

How matter fermions behave under this group? 
The assignment is done in a visibly
asymmetric manner (see table \ref{tab:1}), 
since the ``left'' and ``right'' spinors
have different behaviours (even worser:
only the ``left'' neutrino is 
assumed to exist--an ontological asymmetry).
With the adjectives ``left'' and ``right'', 
we refer to the most elementary objects 
that describe particles with spin 1/2,
called ({\sc Weyl}) 2-spinors\footnote{In fact, 
the Lorentz group SO(1,3) can be seen 
as a  ``complexification'' 
of the group SO(4)$\sim$SU(2)$_L\times$SU(2)$_R,$ 
(the symbol ``$\sim$'' means that the two algebr\ae\ are the same, and
the indices ``$L$'' and ``$R$'' distinguish the two copies of SU(2)). 
We perceive then the close analogy with the usual group of 
rotations, SO(3)$\sim$SU(2) (relation familiar from quantum mechanics).
The representations of the Lorentz group can be then
labelled by a pair of integers $|n,m|;$ 
the smallest non-trivial ones, $|1,0|$ and $|0,1|,$ 
denote left and right 2-spinors respectively. 
Both of them are needed to construct a {\sc Dirac} spinor.}. 
The ``left-right'' asymmetric assignment of 
the elementary particles in the SM accounts for 
a peculiarity of the weak interactions, the violation of 
the parity reflection symmetry.
The generator of electric charge is
the sum of the third SU(2)$_L$ generator 
and of the U(1)$_Y$ (hypercharge) generator $Y$:
$$
Q=T_{3L} + Y
$$
By using table \ref{tab:1} and previous equation, 
one can check that left and right states couple 
with the same charge $Q$ to the photon,
as it should be for parity to be conserved in 
electromagnetic interactions
(the same is true  for the gluons).
At this stage, we introduced 
3 gauge couplings $g_3,$ $g_2$ and $g_1;$
one parameter each gauge subgroup.

\begin{figure}[tb]
\begin{center}
\epsfig{file=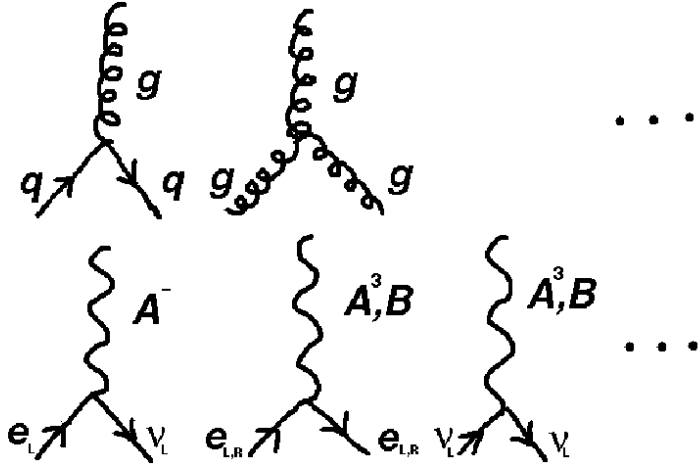,width=10cm}
\end{center}
\vskip-1cm
\caption{Some {\sc Feynman} 
rules (=elementary interactions) of the 
``gauge sector'': the 
quark-gluon coupling, the 3-gluon coupling, and the 
electroweak couplings of the leptons: charged, 
and neutral currents. 
Note the arrows on quark and lepton lines 
(continuous ones), 
that correspond to the ``flow'' of 
U(1)$_Y$ charges. 
Also, left fermions remain left, 
and right ones remain right;
this is a consequence of the 
spin 1 nature of the gauge bosons (wavy lines)
\label{fig:gc}} 
\end{figure}

The elementary interactions 
can be conveniently represented as shown 
in fig.\ \ref{fig:gc}.
Each of these plots corresponds 
to a certain ``{\sc Feynman} rule'' (elementary diagram),
that are used to calculate the 
amplitudes of probabilities of 
the admitted physical processes.
For instance, the {\sc Feynman} 
rule of the first plot (quark-quark-gluon)
tells us the presence of a $g_3$ coupling constant, and 
other factors related to the spin structure, too; 
it corresponds actually to the term of interaction
between $\psi_a$ and $A_\mu^B$ 
in equation\footnote{Usually, a 
theory of the interactions 
is not formulated in terms of  
the equations of motions,
but in terms of a {\em Lagrangian}, 
invariant under the symmetry. 
In this example,
the invariant terms that is represented by the first {\sc Feynman} rule is 
$g T^B_{ab}\times \overline{\psi}_a \gamma^\mu A_\mu^B \psi_b.$
Trying to resemble the notation of section \ref{sec:gp}
(and what is shown in the {\sc Feynman} rule) we can  
illustrate this as an ``elementary reaction''
$\psi_b\to \psi_a A_\mu^B;$ we warn however that
this cannot be considered as an actual reaction
(due to energy non-conservation) but only 
as {\em an element} of a reaction.} 
\ref{exex}.
Joining the elementary 
diagrams in the admitted manners (matching the type of lines)
one obtains the list of the possible 
processes in terms of ``{\sc Feynman} diagrams''.
{\sc Feynman} diagrams are in fact a 
very convenient manner to organise 
the actual computation; 
in the following, however, 
we will use them mostly to illustrate 
the content of the theory, and not to perform computations.

All this story of symmetries and interactions is beautiful, 
but there is an high price to pay for that:
\begin{quote}
{\it No mass is allowed, exactly due 
to $G_{321}$ (standard model) gauge 
invariance!}
\end{quote}

\begin{figure}[tb]
\begin{center}
\epsfig{file=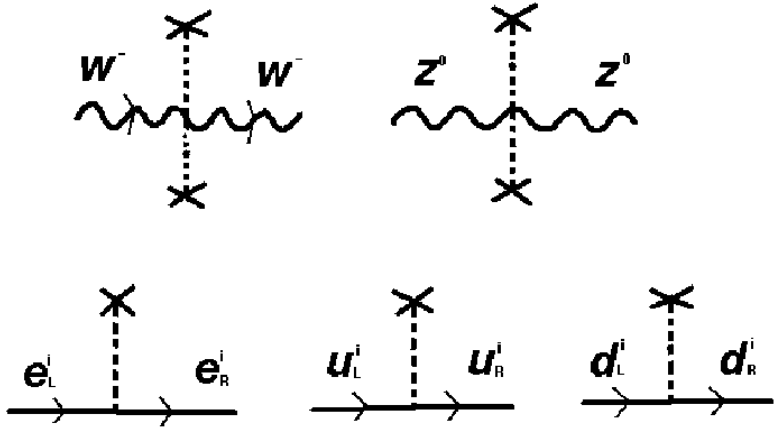,width=11.5cm}
\end{center}
\vskip-2cm
\caption{The {\sc Higgs} particle  
(dotted line) couples with ordinary particles,
and its vacuum expectation value 
(cross at the end of the dotted line) 
breaks the electroweak symmetry and generates mass terms.
The $g_1$ and $g_2$ gauge couplings are 
in the vertices with $W$ and $Z;$ 
the {\sc Yukawa} couplings sit in those 
of quarks and leptons ($i=1,2,3$ is the family index).
Here, the arrows illustrate the ``flow'' of 
electromagnetic charges, that is conserved;
instead, the SU(2)$_L$ or U(1)$_Y$ charges 
flow into the {\sc Higgs} boson and gets lost in the vacuum.
Similarly, the {\sc Higgs} particle
gets mass via its self-coupling. 
\label{fig:yc}} 
\end{figure}
\subsection{Origin of the masses}
\begin{figure}[tb]
\begin{center}
\epsfig{file=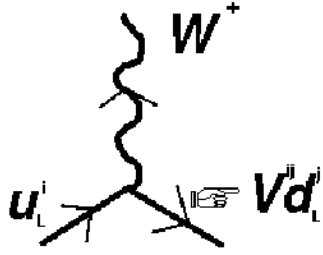,width=5cm}
\end{center}
\vskip-.4cm
\caption{How the charged currents in the SM 
permit the communication between different families. 
The symbol indicated by $V$ in the figure is
an unitary matrix (named CKM matrix 
after {\sc Cabibbo, Kobayashi} and {\sc Maskawa} \cite{ckm}) 
that contains 4 physical parameters:
three angles of mixing (like the three {\sc Euler} angles
of an orthogonal matrix) 
and one complex phase; this latter is the seed of 
CP-violating phenomena.
This delicate sector of the SM will be tested in detail 
by experiments with bottom- and strange-hadrons
(see lecture of {\sc Costantini} at this Conference)
\label{fig:ckm}} 
\end{figure}

Here we come to the second foundation 
of the SM.
The masses arise in a completely 
particular manner: through a {\em symmetry breaking 
ascribed to the vacuum} \cite{higgs}. 
More specifically, it is postulated
that a scalar SU(2)$_L$-doublet exists (the {\sc Higgs} boson), 
with hypercharge $-1/2,$ and it obtains a 
vacuum expectation value:
$$
H=\left| \begin{array}{c} H^0 \\ H^- \end{array} \right|\ \ \ \
\mbox{ such that } {\rm Re}[H^0]\neq 0
$$
This hypothesis is not as strange as it might sound, since
it is a rather common situation in physics of condensed matter, 
{\em e.g.}\ a spontaneous magnetisation breaks
the rotational symmetry\footnote{A similar 
effect exists for strong interactions: 
The vacuum is responsible of the fact that 
the quarks in the proton have an effective 
(``constituent'') mass, and this is
in strict correspondence  with a ``dynamical'' 
reduction of symmetry  \cite{nambu}.}.  
In consequence of this assumption, the electroweak 
invariance is lost (only the electric charge symmetry 
remains untouched), and the $W^\pm$ 
and $Z^0$ bosons acquire mass.
Also, by introducing 9 new parameters between the {\sc Higgs} 
boson and the 
quarks and leptons ({\sc Yukawa} couplings), 
one can account for their masses.
See figure \ref{fig:yc}, and note the
left-right structure of 
the couplings of quarks and leptons
({\sc Dirac} type mass, like the one in section \ref{sec:gp}).
The only communication between different families
in the SM arises at this point, and involves {\em only}
the charged weak currents, To explain this well,
it is necessary to go into some  subtleties.
(1) The particles that have interactions with the $W$ are only 
the doublets, $q_L$ and $l_L,$ so we have interaction 
structures like $u_L^i \to W^+ d_L^i;$
however, (2) (as we see from fig.\ \ref{fig:yc})
the particles with definite mass require 
to match left and right quark (lepton) states, 
namely, $d^i_L$ should match with $d^i_R,$ 
and similarly for $u$ quarks and charged leptons. 
So, these two are different  prescriptions,
or in other terms there is a clash between
the ``interaction eigenstates'' and ``mass eigenstates''.
Normally, people refer to mass eigenstates 
({\em e.g.} the top; the 
electron; {\em etc.}); 
thence charged currents {\em must} be non-diagonal 
(see fig.\ \ref{fig:ckm}). 
For leptons, the absence of neutrino masses 
implies that the non-diagonality is just
formal (of no physical significance): {\em e.g.} $\nu_{\rm e}$ is by definition
the state associated to the electron by charged weak currents.

We are close to the end of the 
list of parameters, and we must introduce now the 
self-interaction of the {\sc Higgs} particle
(scalar potential): 
\begin{equation}
U=\frac{\lambda}{4}\left(|H|^2-v^2 \right)^2
\label{ooqqoo}
\end{equation}
with two parameters, $\lambda$ (the {\sc Higgs} boson
self-coupling)
and $v$ (the vacuum expectation value, 
as it should be clear). 
Notice that we omitted to list an 
additive constant; this 
has no dynamical meaning 
if gravity is ignored, 
otherwise, it should be 
identified as (a contribution to) 
the vacuum energy, and we know by sure
that this is quite small in comparison 
with the ``natural'' scale 
$v^4\approx 10^{50}$ GeV/cm$^3$!!!
Actually, there are still some 
more parameters,
related to the non-abelian gauge bosons 
(one for SU(3)$_c$ and one for SU(2)$_L$)
that formally can be written as interaction terms like
$\theta\times G_{ab} G_{cd} \epsilon^{abcd}.$
While the one related to SU(2)$_L$ is considered harmless, 
the one related to SU(3)$_c$ poses serious problems 
for phenomenology (``strong CP'' problem \cite{theta}), 
and thence has to be small.
We will not list these in the following, 
but we have to stress that these are very 
delicate and mysterious points of the SM; 
surely they indicate (some of) its frontiers.

\subsection{Successes and troubles}

Let us summarise 
the content of the model by some counting. 
The number of matter fermions for family 
is 15,  3 leptons, and 4 quarks which come 
in three types (``colour''). Three families, 45 matter particles. 
Then there are 9 massless spin 1 bosons
(photon and gluons), and three that are massive instead.
The number of parameters is  
$$
18\ (=3\ +\ 9\ +\ 4\ +\ 2).
$$
These are all the fundamental parameters\footnote{We 
put aside those parameters like those in
form factors, partonic distributions, fragmentation functions, 
masses of hadrons... that cannot be calculated and 
have to be measured at present, but that are believed 
not to have a fundamental nature. 
Progresses are expected from computer simulations 
of the SU(3)$_c$ theory on a 
discretized space-time ({\em lattice}).} 
that are compatible with the requisite of 
``renormalizability''.  
In short, this means ``calculability'' of 
the theory; in more diffuse terms, 
a theory of quantised fields (that represent particles) is termed
renormalizable when the quantum fluctuations 
(referred as ``loops'', in connection with their 
representations by {\sc Feynman} diagrams)
produce relatively harmless infinities, namely only those  
that can be re-absorbed in the parameters of the theory. 
So a non-renormalizable theory must be modified 
to become consistent, for instance: \\
(1) adding new parameters (no need 
in the SM--well, 
apart from three more of them, 
the $\theta$ parameters and the cosmological constant,
that we brutally set to zero);\\
(2) or new fields ({\em e.g.}\ if only left electrons were to exist,
the electromagnetic theory would be inconsistent due to 
so-called quantum anomalies; a right handed electron cures
this problem. More complex is the case of 
electroweak interactions, but due
to a conspiracy of all the particles in a family, no such 
trouble exists); \\
(3) or sometimes, it is necessary to reinterpret 
the theory as a non-fundamental one
({\em e.g.}\ effective 4 fermion 
interactions  result when a virtual 
$W$ boson is exchanged between 
two pairs of them; but a 4 fermion 
fundamental interaction, in itself,
would be {\em not} renormalizable).\\
The proof that the SM is 
renormalizable is not simple, however, 
gauge invariance turns out to be a crucial 
ingredient for that \cite{SMren}.
\begin{figure}
\begin{center}
\epsfig{file=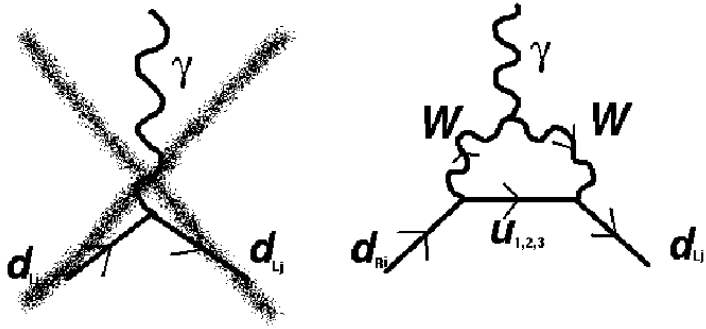,width=10cm}
\end{center}
\vskip-.5cm
\caption{Neither the photon nor the $Z$ bosons
are able to change the family ($i\neq j$) 
at a fundamental level: the first plot--{\sc Feynman} rule--does 
not exist in the SM. 
However this happens at an {\em effective} level:
second plot and other loop diagrams. \label{fig:fcnc}}
\end{figure}

The list of successes 
of the SM is impressivly
long; broadly, can be
divided in those related to 
strong interactions (with few parameters)
and those of electroweak 
interactions (with several parameters).
We will limit here to {\em two} of them, 
that are of rather non-trivial nature
(for a full account, books and reviews 
should be consulted):\\
{\bf  (1) Electroweak precision measurements}\\
Due to the assumption that the {\sc Higgs} particle is a doublet,
the scalar potential has a special structure,
that ensures a relation 
between the $W$ and $Z$  masses and the gauge couplings 
of the SU(2)$_L$ and U(1)$_Y$ groups ($g_2$ and $g_1$) at the 
lowest order:
\begin{equation}
\frac{M_W^2}{M_Z^2} = \frac{1}{1+(g_1/g_2)^2}
\end{equation}
this relation would not hold, 
for instance, if the {\sc Higgs} boson
were in the ``triplet'' representations of SU(2)$_L.$
This argument suggests that 
the hypothesis of ``doublets'' 
is correct, at least in first approximation.
Actually, there are calculable
corrections to this relation 
due to ``loop'' effects, which increase
the left-hand side by:
\begin{equation}
\rho=1+\frac{3}{4} \frac{g_2^2}{(4\pi)^2} \times \frac{m_{top}^2}{M_W^2}
+\mbox{smaller terms;} 
\label{eq:rho}
\end{equation}
$(4\pi)^2$ is the typical loop factor. 
These calculations have been 
first made by {\sc Veltman} \cite{velt}; the 
predictions have been verified 
at {\sc LEP} and other experiments 
(see for instance 
the reviews in \cite{pdg}).
What we want to emphasise is 
that a new parameter, the top mass, 
enters into the game, so that precise measurements of 
$M_W,$ $M_Z$ and of the coupling constants give
informations on the top mass, {\em even if we were to ignore
that the top quark existed\ }\footnote{An analogy can be drawn
with ordinary quantum mechanics: the  
second-order perturbative corrections to the $i$-th level,
$\delta E^{(2)}_i = \sum_j |V_{ij}|^2/(E_i-E_j)$ depends  on 
the intermediate $j$-th levels, even if we don't see 
them directly.}. Similarly, by this method it is possible
to get some information on the {\sc Higgs} boson mass. The sensitivity is
only logarithmic, much weaker than the one to the top mass, 
however present data are precise enough to
show an indication for a 
relatively light {\sc Higgs} boson
(assuming that this is the only new
particle that contributes).\\
{\bf (2) Neutral currents}\\
As we saw, in the SM it is possible to 
change the family only at the price of 
emitting a charged spin 1 boson (fig.\ \ref{fig:ckm}).
However, by a combination of two charged currents,
it is possible to have an {\em effective} neutral 
current transition with change of family. 
The diagram shown in fig.\ \ref{fig:fcnc}
gives in fact an amplitude for electromagnetic 
dipole transition, with a dipole $D$
of the size 
\begin{equation}
D(d_i\to d_j\gamma) = 
V_{ki}^* V_{kj} \times
\frac{e g_2^2}{(4\pi)^2}\times
\frac{m_{d_i}}{M_W^2} \times
f\!\!\left(\frac{m_{u_k}^2}{M_W^2} \right)\ \ \ \ \mbox{where }i\neq j
\end{equation}
($f$ is an adimensional loop function \cite{lf}). 
Beside the crucial electroweak ingredient, 
to obtain the correct quantitative estimate it is necessary 
to consider gluonic corrections \cite{bsg}; so,
it is rather remarkable such a transition
($b\to s\gamma$) has been observed, 
and with a rate that agrees 
with the one predicted in the SM\footnote{There 
is nothing similar in the standard
model for leptonic transition. 
So, a positive observation of, 
say, $\mu\to {\rm e}\gamma$ transition
would be of enormous importance. 
The existence of this
transition is actually predicted in extensions of the SM,
like in simple supersymmetric models.}.

The SM has also some weak points;
we saw the large number of parameters, the
undetected {\sc Higgs} particle 
(that will be discussed further later on),
and the replica of the fermions in three families 
should be added to the list.
But the {\em fatal} failure seems 
to be the following one: 
Neutrinos are predicted to be massless, while 
experiments operated in underground  sites
suggest that they are massive \cite{nuweb}.
This rigidity of the SM is due to 
the global symmetries it has\footnote{To be fair, 
one must remind that ``quantum anomalies'' exist, 
implying that $B+L$ symmetry 
is broken {\em in the SM} \cite{thooft_anom}
($B+L$ is not a gauge symmetry, however!). 
This breaking is manifest when the 
symmetry SU(2)$_L$ gets restored, 
and this has interesting cosmological consequences
({\em e.g.,}\ a  
pre-existing leptonic asymmetry can be converted
into the observed baryonic asymmetry). A more detailed 
discussion is in the contribution of {\sc Auriemma}.}:
A strict conservation of the 
total number of quarks (equivalent to 
the ``baryon number'' $B$); and also of the
number of leptons of each family 
(conservation of the ``family lepton numbers'' 
$L_{{\rm e},\mu,\tau},$
the sum being called {\em the} lepton number $L$). 
In the language of {\sc Feynman} diagrams, 
this means that quark lines are not turned into lepton lines
(or {\em viceversa}); and that the arrows in 
the fermionic lines never clash.
The problem of massive neutrinos is very urgent in our 
view, and will be discussed further in the following.

However, another very evident 
(and, perhaps, deeper)
limitation of the standard model 
of elementary particles is that 
gravity is not included. 
This has to do with 
the fact that quantising gravity 
turned out to be a very difficult problem.
Anyhow, it is rather reasonable to think that 
the model of elementary particle would 
look rather different at the Planck 
scale $(\hbar c/G_{Newton})^{1/2}\approx 1.2\times 10^{19}$ GeV.
There are ideas on how to 
search for manifestations of 
quantum gravity by using ultra-high 
energy cosmic rays\footnote{This
was discussed by {\sc Grillo} in 
this Conference, in connection with the absence of 
a cutoff in cosmic ray proton 
spectrum above $E>5\times 10^{19}$ eV,
which could be related with the 
$\gamma$'s with $E>10^{13}$ eV 
coming from cosmological distances.
See also the 
contributions of {\sc Blasi, Petrukhin} and 
{\sc Stanev.} 
Note that the center of mass energies 
($\sqrt{s}=(2 M_p E)^{1/2}$)  
for cosmic ray proton-proton collisions are 2 TeV 
at the ``knee'', 70 TeV at the ``ankle'' 
and 200 TeV for UHECR's: All 
three exceed the energies of present accelerators.}; 
also, one imagines that a proper 
understanding of black holes would require to make further
steps toward a full theory of gravity.
The concerns regarding the status of the 
cosmological constant stem from 
similar considerations.

\section{Modifying the standard model}

In the following,  we focus 
on massive neutrinos \cite{nuweb}, 
in consideration of 
the urgent character of the problem 
they pose for the SM, and 
also because certain reasonable answers can be offered.
After a first harsh (``phenomenological'') 
approach, we will consider more refined and satisfying 
(``theoretical'') answers; 
ideas like existence of 
``right-handed'' neutrinos, 
``quark-lepton symmetry'' 
and ``grand unification'' 
will be introduced. 
At the end, we will come back to 
{\sc Higgs} particle and to related arguments for 
supersymmetry at the weak scale.

\subsection{A harsh way to massive neutrinos} 
It is not possible to introduce 
masses of neutrinos in the SM. 
However, it is possible to write 
neutrino masses with 2-spinors only;
instead than the usual structure $q_R \to q_L,$ 
one can use  $\overline{\nu_L} \to \nu_L,$ in 
the sense that the role of the 
right state can be played by the {\em anti}-neutrino:
This is what is called {\sc Majorana} mass.
(For reappraisal, one sees that this step is forbidden, unless
one admits that the symmetry U(1)$_Y$ is violated, since
particles and anti-particles have opposite charge---with 
{\sc Feynman} diagrams, one says that this 
mass term produces clashing arrows).
Here, we  postpone the problem of theoretical 
justification, and simply assume the existence of
{\sc Majorana} neutrino mass terms.
These  can be arranged in a mass matrix ${\cal M},$ 
which is symmetric, and thence can be decomposed as follows:
\begin{equation}
{\cal M}_{\ell\ell'}=
\sum_{j=1,2,3} U_{\ell j}^*\ m_j\ \exp(i\xi_j)\ U_{\ell'j}^*
\ \ \ \ \ell,\ell'={\rm e},\mu,\tau
\end{equation}
$m_j$ are the three neutrino masses; 
$U_{\ell j}$ is the MNS 
(after {\sc Maki, Nakagawa} and {\sc Sakata} \cite{mns})
unitary mixing matrix, analogous to the CKM matrix
(and with the same number of physical parameters);
and $\xi_j$ are the so-called ``{\sc Majorana}'' phases
(two of them having physical meaning). All in all,
we have 9 new physical parameters.

\subsubsection{What do we know on massive neutrinos?}
Now we have to face the  question; What is actually 
known from neutrino experiments? 
As reviewed by {\sc Kajita} and {\sc Stanev}
in this conference, there is 
evidence that {\em neutrinos oscillate}, as 
suggested by {\sc Pontecorvo} \cite{pontecorvo}. 
The relevant experiments can be conceptualised
in three main steps; the {\em production} of a neutrino state,
its {\em propagation}, and finally its {\em detection}. 
The states produced and detected are the ``interaction eigenstates'',
that however need not coincide with the ``mass'', or more in general 
with the ``propagation eigenstates''.
In more precise terms, a neutrino 
generated by weak interactions
(say, a $\nu_\mu,$ tagged by a $\mu$ produced in association) 
is supposed to be 
a superposition of propagation 
eigenstates, $\nu_{1,2,3}$; 
if the masses of neutrinos are different, 
these states propagate differently; 
thence, a subsequent detector
could reveal a {\em disappearance} of the 
original neutrino $\nu_\mu$ 
(or the {\em appearance} of a new type neutrino, 
say $\nu_\tau$). This is described by {\em probabilities}
of survival (or of conversion) whose entities
depend on the neutrino 
mixing and the differences of masses squared.
So: (1) atmospheric neutrinos 
inform us on one squared
mass difference and one 
mixing angle (these results 
can be regarded as indication of 
$\nu_\mu$ disappearance; {\sc Super-Kamiokande} gives also
indications of $\nu_\tau$ 
appearance {\em via} 
neutral currents events); 
(2) solar neutrinos 
inform us on another (squared)
mass difference and another 
mixing angle\footnote{In fact, 
more than one region of the 
parameter space is compatible
with present information. 
This indication of massive neutrinos
will largely deserve further studies 
and attentions.}  
(this results can be regarded as 
$\nu_{\rm e}$ disappearance; neutral 
current signals can 
be seen by {\sc Super-Kamiokande} 
and {\sc SNO} detectors {\em together}).
So, there are 5 parameters that are simply unknown at present.
They are\footnote{In this context, the LSND indication 
cannot be explained. So, this might open further space 
for surprises.}
\begin{enumerate}
\item the mixing between $\nu_{\rm e}$ and
the state responsible of atmospheric oscillations, $|U_{{\rm e}3}|;$
\item a phase that can cause CP violating oscillations (included 
in $U_{\ell i}$);
\item the mass of the lightest neutrino;
\item the {\sc Majorana} phases $\xi_j.$
\end{enumerate}
We have plenty of choice of what to study next! 
However, difficulties are not lacking, too...

\subsubsection{Can we measure something {\em more}?}
Let us discuss the parameters of the previous list.

{\em (1)} The first parameter 
is bound by the {\sc CHOOZ} experiment to be small, 
$|U_{{\rm e}3}|^2<2.5\times 10^{-2}$ 
(actually, the bound depends to a 
certain extent on the mass differences).
In fact, there are some 
theoretical arguments\footnote{Theoretical 
expectations are not misleading, if considered
in the proper manner; however, we feel 
necessary to stress at this point 
the provisional character of 
these expectations.} 
suggesting  that this parameter is even smaller,
at the level of $|U_{{\rm e}3}|^2\sim 10^{-3};$
these will be discussed in section \ref{sec:ql}.
So, the effects on oscillation probabilities 
could be quite tiny
and would be not possibly found even after refining 
cosmic ray neutrino experiments. 
Actually, it seems rather 
difficult that present generation 
of long-baseline experiments will achieve 
sensitivity to such a small value 
(compare with the report of {\sc Scapparone} for this Conference).
A completely different 
possibilities is offered 
by the detection of 
a type-II supernova, 
since the neutrino burst are
substantially affected by 
the presence of even such a small value of $|U_{{\rm e}3}|$ 
\cite{alyosha}.

{\em (2)} As for the CP violating phase, 
it is rather difficult to foresee the perspectives of measurement
at the present stage. It seems to us that success could
be possible only if the solar neutrino problem has
relatively large values of the mass difference
(``large angle {\sc Mikheyev-Smirnov-Wolfenstein}'' 
solution \cite{msw}--see contribution of {\sc Kajita}); 
this possibility could be tested
by the {\sc KamLAND} experiment which aims at 
detecting neutrinos from distant reactors.

{\em (3)} The most sensitive tool to the 
lightest neutrino mass is 
the search of anomalies in the endpoint 
spectra of $\beta$ (=electron)
emitters with low $Q$-value, 
since in that kinematical regime the mass 
of the neutrino emitted in association 
becomes relevant. 
The experimental searches ({\sc Troitsk, Mainz})
bound the mass of this neutrino  
to be below the few eV range
(theoretically, it is not excluded that the
actual value is close to this one; 
in that case, however, oscillation experiments
would force the conclusion that the spectrum is almost degenerate,
which does not sound particularly appealing).
Note that close type-II supernovas could
give some information on this parameter, too. 
Also, cosmological considerations give valuable information
on massive neutrinos as sub-dominant components of the Universe
(assuming of course that the main 
components and the dynamics are
well understood), since the big-bang 
picture suggests the existence of a sea 
of {\em relic neutrinos}, a close relative of the cosmic 
microwave (photon) background. In passing, we have to say
that it is very hard to conceive other (non-gravitational)
tests of the {\em relic neutrinos} hypothesis.

\begin{figure}
\begin{center}
\epsfig{file=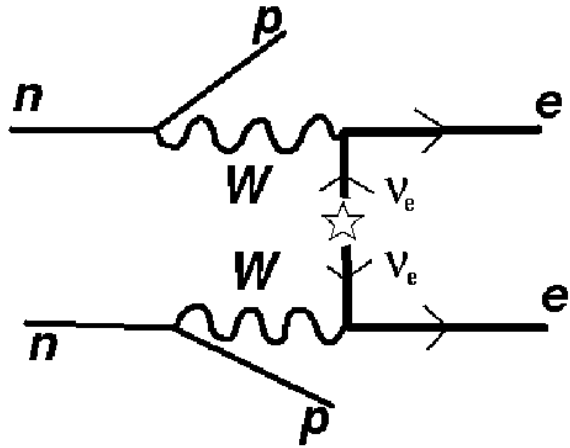,width=8.5cm}
\end{center}
\vskip-1cm
\caption{Diagram for neutrinoless-$2\beta$
decay at level of nucleons ($n,$ $p$ are neutrons and protons).
Note the clashing arrows in the the neutrino lines,
that are characteristic of the {\sc Majorana} mass of neutrinos,
and signal the violation of the lepton number. \label{fig:nlbd}}
\end{figure}

{\em (4)} The {\sc Majorana} phases 
are very-very difficult to measure directly. 
Actually, the only handle of which 
we are aware is a rather indirect one, namely 
a measurement of these parameters in combination 
with other ones, which could be possible 
observing the neutrinoless-$2\beta$ 
decay (see fig.\ \ref{fig:nlbd}). 
In this transition, a nucleus increases its charge
by two unities, by emitting two electrons simultaneously.
If detected, this transition could 
inform us on the size of $|{\cal M}_{\rm ee}|^2,$ which depends
{\em also} on the {\sc Majorana} phases.
The uncertainties in the description of nuclear effects
could limit quantitatively the precision of our inferences; however,
a detection of this transition  
would be of great significance, since it would strongly suggest that 
neutrino have {\sc Majorana} type mass: oscillation experiments
cannot help do this. The present limit on $|{\cal M}_{\rm ee}|$
has been obtained by the {\sc Heidelberg-Moscow} 
Collaboration and is in the $0.1-1$ eV range.

However, even in the optimistic assumption 
all the unknown quantities we discussed will be measured,
it is evident that one of the parameters of the neutrino mass 
matrix will remain \underline{unmeasured} ($5-4=1$).

\subsection{A kind way to massive neutrinos}
Now, we discuss some (theoretically) respectable extensions
of the standard model, with massive neutrinos
and in general with new physics.

\subsubsection{Adding new particles}
\begin{figure}
\begin{center}
\epsfig{file=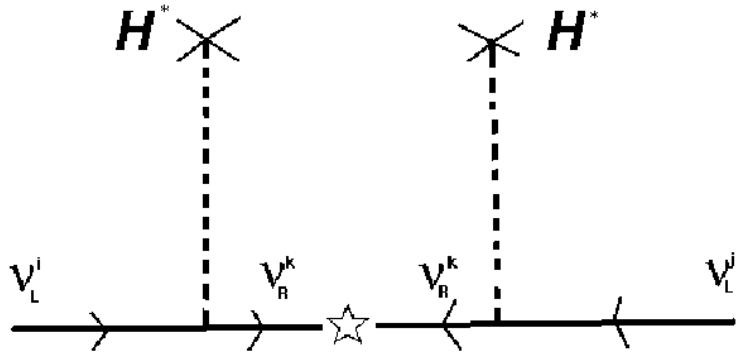,width=11cm}
\end{center}
\vskip-0.7cm
\caption{Diagram for the see-saw neutrino masses.
The right-handed neutrinos are supposed to be heavy, 
and act only as quantum fluctuations. However, they 
introduce a new ingredient, the violation of the lepton number
(testified by the clashing arrows). Instead, the U(1)$_Y$ charge
flows in the vacuum through the {\sc Higgs} boson
lines. \label{fig:ss}}
\end{figure}

We start considering the {\sc Higgs} boson $H$ and 
the left-leptons $l_L$, that have the same gauge numbers, 
$|1,2,-1/2|$ (notation as in table \ref{tab:1}).
An admitted interaction must be invariant, 
so the product $l_L H^*$ seems to be OK. 
But it is the space-time structure that is not OK, since 
combining a spinor and a scalar we do not obtain an 
invariant. One simple solution is to introduce 
a new fermion $\nu_R,$ with no gauge numbers, and construct
$l_L H^* \overline{\nu_R}.$ 
The answer we get is rather obvious\footnote{This procedure, 
however, is of wide validity \cite{weinberg_opp}.
If for instance we wonder which particle 
couples a lepton and a quark, we have just to multiply two such 
fermions, and deduce from gauge invariance the properties
of this hypothetical particle. 
Also, we can list the {\em effective} interactions
that are gauge invariant, which do not satisfy the
requisite of renormalizability (``higher dimensional'' 
operators).}:
we just introduced something 
that must be called ``right-handed'' neutrino,
and we obtained 
a {\sc Yukawa} interaction completely analogous 
with quark or charged leptons {\sc Yukawa} terms
(this term alone would 
originate what is called 
a {\sc Dirac} mass for neutrinos). 
But we learned that $\nu_R$ has no gauge interactions,
so for consistence we must consider the invariant 
$\nu_R\nu_R,$ too (={\sc Majorana} mass for the right-handed neutrino).
These two elements give rise to the famous ``see-saw'' model for
the masses of the left-neutrinos \cite{seesaw}
(which is {\sc Majorana} in type;
the name comes from the fact that, the more the 
right-handed neutrino masses go up, the 
more the left-neutrino masses go down): 
see fig.\ \ref{fig:ss}.
It should be noted that, even if the 
right-handed neutrino has super-heavy mass and cannot be 
produced at accelerators, it still leaves
its footprints in the masses of left-neutrinos.
The generated mass can 
be regarded {\em effectively} as a $(l_L H^*)^2$ term;
it reveals its non-fundamental nature 
from the fact that it is not renormalizable
(one can draw here a very close analogy with
the 4 fermion weak interactions).
Note that the number of parameters is still enlarged!
However, some of these parameters might have some 
interesting manifestations,
for instance they can generate a leptonic asymmetry
in the early Universe \cite{fy}.

There is at least one alternative possibility 
that we feel should be mentioned: if, in 
order to obtain {\sc Majorana} masses
for neutrinos,  we directly 
multiply two leptonic 
doublets, we find the quantum numbers
$|1,3+1,-1|.$ 
Thence, by postulating a triplet scalar $\Delta$ with 
U(1)$_Y$ charge $-1,$ we can construct the invariant term 
$l_L l_L \Delta;$ and if we give a {\em tiny} 
(see eq.\ \ref{eq:rho} and discussion therein)
vacuum 
expectation value to the neutral component of the triplet, we 
find again {\sc Majorana} masses.
So, the question arises, which 
of these (or other) mechanism for massive 
neutrinos is the correct one? 
This is very difficult to answer convincingly, and it could
be considered a frontier of the ``theory beyond the SM''!

\subsubsection{Larger gauge groups}
We proceed in the presentation of promising  ideas, and 
pass to the concept of {\em grand unification} 
\cite{gg_ps_gfm}.
The existence of right-handed neutrinos--whose interest 
has been already discussed--can 
be argued on the basis of quark-lepton symmetric spectrum
(even if the {\sc Majorana} mass of the right-handed neutrinos
is by itself a point of asymmetry). 
This hypothesis becomes 
compulsory if we assume that the SM is a group of 
residual symmetry of certain ``unification groups.''
We just present few examples 
and limit ourselves to illustrate how the matter fermions 
(and the right-handed neutrinos)
fit in the representations of larger groups:\\[3ex]
\noindent
{\large \bf SU(5)}. ({\sc Georgi-Glashow}) This group of $5\times 5$ 
matrices obviously includes 
SU(3)$_c$ and SU(2)$_L$ as the upper $3\times 3$ and lower 
$2\times 2$ blocks, while U(1)$_Y$ is in the
diagonal. The representations where the matter fermions sit are:
\begin{equation}
{\bf 10=}\left| \begin{array}{ccccc} 
0 & \overline{u_{R3}} & -\overline{u_{R2}} & {u_{L1}} & d_{L1} \\
  & 0  & \overline{u_{R1}} & {u_{L2}} & d_{L2} \\
  && 0 &   {u_{L3}} & d_{L3} \\
  &&& 0 & \overline{ {\rm e}_R} \\
  &&&& 0 
\end{array}\right|\ \ \ \ \
{\bf \bar{5}=}\left| \begin{array}{c} 
\overline{d_{R1}}\\ 
\overline{d_{R2}}\\ 
\overline{d_{R3}}\\
\nu_L\\
{\rm e}_L 
\end{array}\right| \ \ \ \ \
{\bf 1=}\ \nu_R
\end{equation}
(1,2,3 are the index of SU(3)$_c;$ 
the {\bf 10} matrix is antisymmetric). 
In this context, the  
right-handed neutrino has  the same 
status as in the SM.\\[3ex]
{\large \bf SU(4)$\times$SU(2)$_L\times$SU(2)$_R$} ({\sc Pati-Salam}). 
This group is very satisfying as for the quark-lepton symmetry.
The fermions are assigned to:
\begin{equation}
{\bf |4,2,1|=}
\left|
\begin{array}{cccc}
u_{L1} & u_{L2} & u_{L3} & \nu_{L} \\
d_{L1} & d_{L2} & d_{L3} & {\rm e}_{L} 
\end{array} 
\right|
\ \ \ \ \ \ 
{\bf |4,1,2|=}
\left|
\begin{array}{cccc}
u_{R1} & u_{R2} & u_{R3} & \nu_{R} \\
d_{R1} & d_{R2} & d_{R3} & {\rm e}_{R} 
\end{array} 
\right|
\end{equation}
(note the presence of right-handed neutrinos).
Leptons, in a sense, are 
just the fourth type of quarks.
For us is difficult to 
see this, and maintain the opinion 
that it has no meaning.
One could consider a weak point of this assumption
the fact  that there are three different 
(non-unified) gauge subgroups (however, 
by imposing a parity that 
relates $L$- and $R$-subgroups this 
improves).\\[3ex]
{\large \bf SO(10)} ({\sc Georgi, Fritzsch-Minkowski}). This group includes
SU(5) (since a complex 5 vector can be 
obviously mapped in a real 10 vector) 
but also the Pati-Salam group (since SO(6)$\sim$SU(4)). 
Each family of  matter fermions sits in a single representation:
\begin{equation}
{\bf 16=10+\bar{5}+1=|4,2,1|+|4,1,2|}
\end{equation}
Like SU(5), this group has an unique coupling. However,
the breaking of SO(10) down to the SM might take place in several
steps (with several ``intermediate'' scales).\vskip.3cm

Whatever the gauge group, it is 
needed that the large gauge 
symmetries are broken (and also that the 
fermion masses are generated).
This calls for scalars, and
opens up many possibilities
(and troubles, see section \ref{sec:troub}). 
We mention only one specific point 
here \cite{goran}.
As we saw, the SO(10) or Pati-Salam group naturally includes
right-handed neutrinos, which is certainly a good thing.
However,  the reduction of SU(2)$_R$ symmetry 
typically requires the 
existence of scalar ``right'' triplets;
the symmetry forces the existence of ``left'' triplets; 
so that, there are two competing sources 
for massive neutrinos: see-saw and ``left'' triplet.

Apart from neutrino masses, 
the unification groups have (some)
predictivity on the 
{\em unification of gauge couplings,} 
{\em of fermion masses,} 
and also on {\em proton decay}. 
From the experimental 
point of view, the last aspect 
is surely the most interesting. 
Proton decay could be 
due to the new gauge bosons
of the grand unified group, that permit 
communication between quarks and leptons;
however there are also other possibilities,
{\em e.g.}\ new scalar particles might 
have an important role (this happens commonly
in supersymmetric models).
Trying to summarise in a few 
words the present situation:
proton decay is not experimentally found 
(strongest limits come from the 
{\sc Super-Kamiokande} experiment)
but there are some theoretical 
models that offer hopes of detection for future 
detectors (new detectors like 
{\sc ICARUS} have superior properties, but
its {\em mass} might  
limit its discovery potential ... 
should a {\sc Mega-Kamiokande} be built?). 
Experimental and theoretical proposals, however, are still at
a stage of discussion; the ``future'' seems not close.

\subsubsection{Use of quark-lepton symmetry \label{sec:ql}}

Here, we present an {\em ansatz} 
for massive neutrinos, that 
we consider quite reasonable
since it is based on a principle: 
quark-lepton symmetry (or more precisely,
quark-lepton correspondence).
One starts to note that 
the masses of up-type quarks increase 
strongly changing family, 
in comparison with what happens for down type quarks,
or also for charged leptons.
This difference of hierarchies 
motivate the assumption that neutrinos 
have still weaker differences among them.
An actual implementation of this idea, due to 
{\sc Sato} and {\sc Yanagida}\footnote{This model 
has been inspired by SU(5), since the weaker hierarchy was explained by
saying that the ${\bf \bar 5}$-plets of second and third 
families (that contain $\mu$ and $\tau$ 
neutrinos) do not pay-off hierarchy factors; while this always happen for
${\bf 10}$-plets.} \cite{sy}, 
is the assumption that the neutrino mass has the structure:
\begin{equation}
{\cal M}\propto
\left|
\begin{array}{ccc}
\varepsilon^2 & \varepsilon & \varepsilon \\
& 1 & 1 \\
&& 1
\end{array}
\right|\ \ \ \ \ \varepsilon=1/20\sim \sin^2\!\theta_C\sim  m_\mu/m_\tau
\end{equation}
(where the {\sc Cabibbo} angle 
$\theta_C$ is a parameter of the CKM matrix).
To make the statement sufficiently 
vague (or equivalently, sufficiently precise) 
one postulates that the elements of 
the matrix include also {\em coefficients order unity}
(these are in fact are essential in order to generate
three different neutrino masses).\\ 
The weak points of this assumption are that:\\ 
(1) the overall scale is not predicted;\\
(2) the hierarchy of masses between the two heavier neutrinos
tends to be rather weak.\\ 
The advantages (after the weak points are made up, 
by adjusting the unknown ``coefficients'' and the overall scale) are that:\\ 
(1) a large mixing for atmospheric neutrinos
is automatic;\\
(2) there is a prediction of ``large angle 
{\sc Mikheyev-Smirnov-Wolfenstein}'' solution 
(with the correct type of mass hierarchy);\\
(3) the value of $U_{{\rm e}3}$ is  predicted to be small.\\
Note, incidentally, that also 
neutrinoless $2\beta$ transitions are 
predicted to be suppressed, 
since $|{\cal M}_{\rm ee}|^2\sim \Delta m^2_{atm} \times 
\epsilon^4.$
This is essentially a manifestation of the fact that neutrinos
are supposed to obey a sort of family hierarchy, and
is tightly related to the suppression of $U_{{\rm e}3},$ since
$U_{{\rm e}3} \sim \epsilon .$

\subsection{Troubles with fundamental scalars, 
and supersymmetry as solution \label{sec:troub}}
\begin{figure}
\begin{center}
\epsfig{file=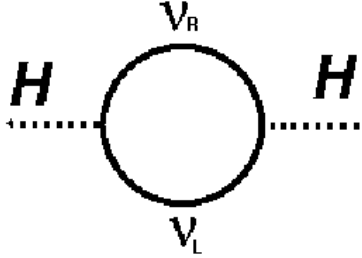,width=6cm}
\end{center}
\vskip-1cm
\caption{An example of quantum communication (loop diagram)
between hypothetical heavy mass scales--right-handed neutrino
in this example--and the electroweak--{\sc Higgs}--mass 
scale. \label{fig:hp}}
\end{figure}

Here we come to one trouble of the SM 
(even more severe in the unified theories).
This is, in essence, a 
{\em problem of hierarchy of scales} \cite{hier}.
One can say that, due to quantum fluctuations, 
heavy masses creep in the {\sc Higgs} boson ($\sim$ weak boson) 
scale and want to destabilise it.
For instance, the diagram in the fig.\ \ref{fig:hp} changes 
the coefficient of the bilinear term 
$\mu^2\times |H|^2$ by an amount of the order of
\begin{equation}
\delta \mu^2 \sim \frac{Y^2}{(4 \pi)^2} \times M_{\nu_R}^2\ \ \ \ 
\mbox{ where }\ \ \ \ 
\left\{
\begin{array}{l}
Y=\mbox{{\sc Yukawa} coupling of neutrinos,} \\
M_{\nu_R}=\mbox{Mass of right-neutrinos.} 
\end{array}
\right.
\label{hhh}
\end{equation}
which (apart for the loop pre-factor)
seems to produce a huge scale, comparable to $M_{\nu_R}$
(not of the order of the electroweak scale as we need). 
To be picky, one can say 
that the individual contributions to $\mu^2$ are
not separately measured; that only their sum 
has physical meaning; and that we should not 
ask the theory to say more than it can.
However, in this manner we 
would most probably give up 
any chance of {\em predicting} 
these fundamental parameters, since a hypothetical 
(more complete) theory would be forced to 
explain a very precise fine-tuning\footnote{Note 
the purely \underline{theoretical} character 
of this problem; in this 
sense, one could say that there 
is a disturbing situation with fundamental scalars, 
but not an untolerable one.}.
This situation motivated several 
extensions of the SM, and all of them
have new physics (close) at the electroweak scale;
for instance, it was postulated 
that the {\sc Higgs} particle 
is not fundamental---but instead a pion-like object
({\em technicolor}).

We will concentrate 
the discussion on
{\em supersymmetric} models \cite{susy}.
Supersymmetry is an extension 
of the space-time group, which relates 
fermions and bosons by a symmetry transformation.
In the models that could be possibly relevant for 
electroweak scale physics, supersymmetry commutes
(=is unrelated) with the gauge group; 
so that any ordinary particle\footnote{In 
principle, the lepton doublet and one 
of the {\sc Higgs} bosons could be ``partners'';
in practise, this type of model would 
have too large violations of the 
leptonic number (and other troubles) 
and for this reason  is not pursued.}
obtains a ``partner'', 
and we have scalar electrons (sleptons), 
fermionic gluon (gluino), {\em etc.} Actually, it is necessary 
that the number of {\sc Higgs} doublets 
is at least {\em two}\footnote{A problem that we cannot 
discuss here is: How the additional scalars 
of the model are prevented from obtaining 
a vacuum expectation value? We have to limit ourselves to say  
that {\em for some value of the parameters} 
it can be done.}.
We recommend to make reference to the contributions 
of {\sc Ganis} and {\sc Denegri} 
for a more detailed description
of these models, and discussion of 
the perspectives of confirmation. 
The connection with the 
``hierarchy problem'' is due to an amazing property of supersymmetry
as a quantum field theory: that the 
``loops'' involving bosons and fermions compensate each other,
and contributions like those in eq.\ \ref{hhh} \underline{do not arise}.
At this point, however, we have to recall that an 
extension of the SM should have
{\em broken} supersymmetry in order to be realistic
(otherwise, for instance, the ``partners'' would have 
the same mass of the ordinary particles). The actual mechanism
for supersymmetry breaking is an  open question at present, 
which surely is not a nice feature, 
even if there are reasonable 	proposals.
However, if one assumes that the breaking scale is order of the TeV,
and restricts the allowed breaking terms to so-called ``soft-breaking'',
the quantum properties are maintained and the hierarchy problem is  
under control\footnote{Despite the desire to mantain as much 
predictivity as we can, we are forced to introduce 
new parameters in order to do this (for a precise counting, see
for instance the review on supersymmetry in \cite{pdg}).}. 
In this paragraph, we presented an 
instant summary of the idea of ``supersymmetry 
at the weak scale''. Much more could be said,
however, the key question  that we have to answer is: 
Do these considerations have any relevance 
to the description of Nature?
Let us list three considerations that 
suggest (but not ``imply'') an affirmative answer:
\begin{figure}[tb]
\begin{center}
\epsfig{file=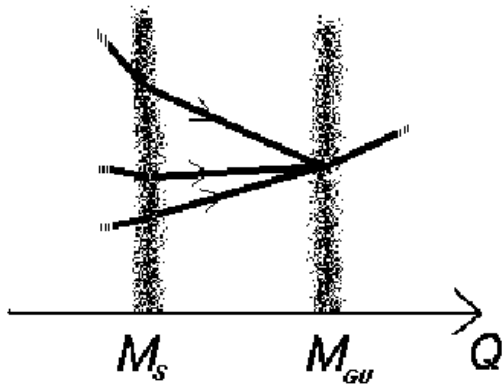,width=7cm}
\end{center}
\vskip-.7cm
\caption{Sketchy plot of the unification
of SM gauge couplings in supersymmetric context.
The two vertical bands  indicate 
where supersymmetry begins
(scale $M_S$)
and where the unified dynamics begins
(scale $M_{GU}$); they are sources of 
uncertainties for the predictions. \label{fig:gu}}
\end{figure} 
\begin{itemize}
\item {\bf Electroweak precision measurements suggest the existence
of a light {\sc Higgs} particle
as predicted in (minimal) supersymmetric extensions
of the standard model}.\\ 
The prediction is mainly 
due to the fact that the 
{\sc Higgs} boson
self-coupling $\lambda$ in eq.\ \ref{ooqqoo}
turns out to be a combination
of the (measured) electroweak 
gauge constants (the scalar potential 
is quite constrained in these models).
This prediction can be tested at future colliders
(... or if we are really lucky, even before; see {\sc Denegri},
these Proceedings). 
\item {\bf Gauge couplings unify in the context of low energy
supersymmetric model.}\\
Here  we mean that: the extrapolation 
of the couplings is compatible with 
the hypothesis of grand unified dynamics (broken
below a certain scale $M_{GU},$ see fig.\ \ref{fig:gu}) \cite{gcu}.
This does not happen in the ordinary SM 
(not within a model with a single scale of breaking).
It is rather remarkable that the large 
unification  scale, $M_{GU}\sim 2\times 10^{16}$ GeV, 
is  comparable with what is suggested by a see-saw mechanism 
for massive neutrinos.
Can proton decay provide the crucial confirmation of this indication? 
Specific channels exist \cite{pd}, like 
\begin{equation}p\to K^+ \bar{\nu};\end{equation} 
the presence of a kaon
(the characteristic aspect)
results  from the fact that proton 
decay is expected to 
be  due to {\sc Yukawa}  type interactions, 
and to observe for this reason the family hierarchy
(see fig.\ \ref{fig:spd}).
However, the dependence on the unknown aspects 
of the model is strong (and no doubt that the usual {\sc Yukawa} sector 
already challenges our understanding), 
and, theoretically, it is not excluded  
that the proton decay process is rather suppressed.
\begin{figure}[tb]
\begin{center}
\epsfig{file=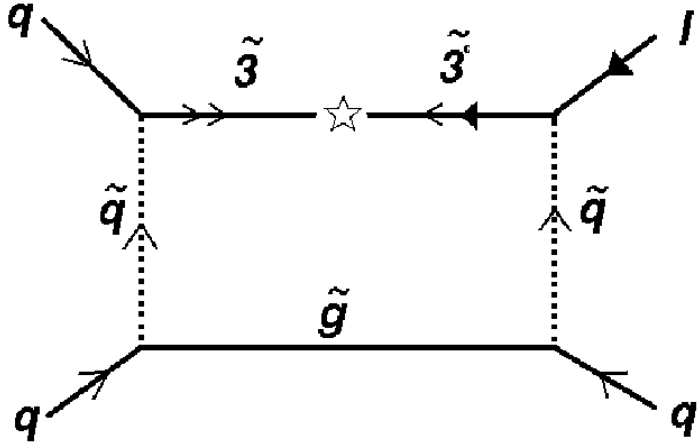,width=10cm}
\end{center}
\vskip-1cm
\caption{One diagram for supersymmetric proton decay.
$\tilde q$ and $\tilde g$  
are the supersymmetric partners
of the quarks and gluons; instead, the 
SU(3)$_c$ triplets $\tilde {\em 3}$ and $\tilde {\em 3}^c$ are 
fermions, that are paired by grand unification 
with the supersymmetric partners of the 
{\sc Higgs} particles. 
The open arrows indicate the 
flow of baryon number, the closed ones 
the flow of lepton number; the star is 
the point of clash. \label{fig:spd}}
\end{figure}
\item {\bf A cold dark matter candidate exists.}\\
The lightest supersymmetric particle (LSP)
is stable due a discrete symmetry that can be 
incorporated in the model (which, for the record, is 
called R parity\footnote{For completeness, 
we must add that theoretical models 
can be constructed in which this symmetry is 
broken; the cold dark matter
candidate disappears, but such a breaking could 
account for massive neutrinos.
At present, however, this possibility is not 
considered of particular appeal, due to the rather
{\em ad-hoc} values of the parameters 
that are required to account for the masses of the neutrinos.}).
Is the galactic ``LSP cloud'' \cite{cdm} 
responsible of the modulated signal seen in 
the {\sc DAMA} experiment shown by {\sc Incichitti} at this 
Conference?
If this were true, this result would be of
enormous importance, not 
only for the direct detection of dark matter
but also as a first signal of ``supersymmetry at the electroweak scale''; 
and it would also open quite interesting 
perspectives for future collider searches, 
as discussed by {\sc Ganis}.
Further studies and confirmations 
are of essential importance ({\sc Incichitti}).
\end{itemize}

We would like to add a 
comment on the {\sc Higgs} boson mass. 
A value on the large side (say, $>120$ GeV)
would indicate in a supersymmetric context
a rather strong hierarchy between 
the vacuum expectation values of the two
{\sc Higgs} doublets, $\langle H_{up}^0\rangle \gg 
\langle H_{down}^0\rangle.$ 
This would suggest an entire series
of theoretical and phenomenological 
questions; for example, the ({\sc Yukawa} type) 
proton decay is expected to be 
enhanced in this regime.
Instead, if the {\sc Higgs} boson mass 
turns out to be really large 
(say, $>150$ GeV)
it seems not easy to avoid the 
conclusion that ``supersymmetry at the 
electroweak scale'' is in trouble; 
this will be the crucial 
test of the model.
Finally, we remind that in 
the SM there is a limit on the {\sc Higgs} boson mass
suggested by the consideration that
the self-coupling 
$\lambda$  should 
be not driven negative as an  effect of 
the quantum fluctuations, 
(``vacuum stability'' \cite{vst}) at least 
up to the Planck scale 
where new effects most probably appear.
It is rather 
funny, but this {\em lower} limit almost coincides with the 
{\em upper} limit in the
(minimal) supersymmetric extension of the standard model. 
So (if a joke is permitted) 
we present a prediction for {\sc LHC}:
\begin{equation}
m_H=135\pm 5 \mbox{\ GeV}
\end{equation}
the reason is that this 
value will increase the 
entropy in the minds of 
several theorists.
Note, however, that the decay of a 
standard and supersymmetric {\sc Higgs} particles 
with the same mass (or also the production 
rate--``cross-sections'')
could be rather different; 
thence, these measurements would  offer 
a possibility to distinguish between 
the SM and its supersymmetric extension 
even in this tricky case.

\section{(Not quite a) conclusion}
We would like to close this pages
by spending few words of caution, to remind
that failures of the standard model 
have been often claimed in the past years
(today, several of 
them are considered 
dubious or simply wrong tracks). 
Here is an arbitrary selection:
$$
\begin{array}{lcl}
\mbox{\sc  THEORETICAL} & &\mbox{\sc EXPERIMENTAL} \\
\mbox{\sc  INTERPRETATION} & &\mbox{\sc ANOMALY} \\ 
& & \\
\mbox{leptoquark} &................................\ \ \ \ \ \ & \mbox{High $x$ and $Q^2$ events at {\sc HERA}}\\
\mbox{compositeness} &................................\ \ \ \ \ \ & \mbox{Excess of 4-jet events at {\sc ALEPH}}\\
\mbox{light gravitino} &................................\ \ \ \ \ \ & \mbox{${\rm ee}\gamma\gamma E\! \!\!\!/$\ \ event at {\sc CDF}}\\
\mbox{17 keV neutrino} &................................\ \ \ \ \ \ & \mbox{bump in $\beta$ spectra ({\sc Simpson, ...})}\\
\mbox{monopole} &................................\ \ \ \ \ \ & \mbox{induced currents ({\sc Cabrera})}\\
\mbox{proton decay} &................................\ \ \ \ \ \ & \mbox{contained multitrack events at {\sc KGF}}\\
\mbox{...} &................................\ \ \ \ \ \ & \mbox{...}
\end{array}
$$
Is there any moral 
behind these stories?
Maybe not; however:\\
1) they suggest to go slowly and carefully 
from data to theories and back 
(because of  possible pitfalls of interpretation, of suggestion, {\em etc.});\\
2) they witness how hard 
is to reach the frontiers 
of standard model; and, also, how 
strong is the desire of particle physicists 
to find them!

\vfill
\noindent\hrulefill

I am grateful to the Organizers 
and Partecipants 
(in particular to
{\sc B.\ Alessandro, 
R.\ Bandiera, 
P.\ Blasi, 
S.\ Colafrancesco, 
F.\ Giovannelli,
A.\ Grillo, 
G.\ Mannocchi, 
R.\ Ramelli, 
P.G.\ Rancoita,
E.\ Scapparone, 
A.\ Stamerra, 
A.\ Surdo}) for 
the most pleasant and 
informative discussions,
and to {\sc F.\ Cavanna} 
for a careful reading 
of the manuscript.
I would like to 
take this occasion
to thank
{\sc R.\ Barbieri, 
V.\ Berezinksy,
S.\ Bertolini, 
W.\ Buchm\"u{}ller,
A.\ Di Giacomo,
A.\ Masiero, 
N.\ Paver,
S.\ Petcov, 
E.\ Roulet, 
G.\ Senjanovi\'c,
A.Yu.\ Smirnov,
M.\ Veltman}
and 
{\sc T.\ Yanagida}
to whom I owe what 
I know on the SM 
and its extensions,
and who largely deserve the 
credit for niceties in the
presentation; errors 
and misinterpretations
of course are mine.

\end{document}